\documentclass[prl,twocolumn,citeautoscript,superscriptaddress,showpacs]{revtex4}

\usepackage[dvips]{graphicx}
\usepackage{epsfig}
\usepackage{dcolumn}

\usepackage{amsmath}

\begin{document}
\title{The absorption spectrum of hydrogenated silicon carbide
  nanocrystals from \emph{ab initio} calculations}

\author{M\'arton V\"or\"os} \affiliation{Department of Atomic Physics, Budapest
  University of Technology and Economics, Budafoki \'ut 8., H-1111, 
  Budapest, Hungary}

\author{P\'eter De\'ak} 
\affiliation{Bremen Center for Computational Materials Science,
               Universit\"at Bremen, Otto-Hahn-Allee 1, 28359 Bremen,
               Germany}

\author{Thomas Frauenheim} 
\affiliation{Bremen Center for Computational Materials Science,
               Universit\"at Bremen, Otto-Hahn-Allee 1, 28359 Bremen,
               Germany}

\author{Adam Gali} 
\affiliation{Department of Physics and School of
  Engineering and Applied Sciences, Harvard University, Cambridge,
               Massachusetts 02138, USA}
\affiliation{Department of Atomic Physics, Budapest
  University of Technology and Economics, Budafoki \'ut 8., H-1111, 
  Budapest, Hungary}

\begin{abstract}
  The electronic structure and absorption spectrum of hydrogenated
  silicon carbide nanocrystals (SiCNC) have been determined by first
  principles calculations. We show that the reconstructed surface can
  significantly change not just the onset of absorption, but the
  \emph{shape} of the spectrum at higher energies. We found that the
  absorption treshold of the reconstructed SiCNs cannot be accurately
  predicted from traditional density functional theory calculations.
\end{abstract}

\maketitle

Semiconductor nanocrystals (NC) are objects of intense interest in
fields ranging from biology \cite{Michalet05} to third generation
solar cells \cite{Gur05,Klimov07}. NCs are small pieces of crystal
with a surface to volume ratio that is much larger than is typical for
bulk crystals. Therefore, understanding the surface states of NCs
could be critical in the interpretation of the measurements, or in
applications \cite{Pandey08}.  Pure covalent semiconductor NCs have
dangling bonds at the surface that can easily react with the molecules
present in the environment. The simplest example is when the dangling
bonds are saturated by hydrogen atoms. It is well-known from surface
studies of bulk covalent semiconductors that saturation can happen
either by conserving the bulk-like structure at the surface or by
reconstruction at the surface with formation of long bonds. In what
follows, we will refer to a non-reconstructed surface as an ``ideal''
surface. While the structure of surface reconstruction and its effect
on the electrical/optical properties can be monitored by experiments
and first principles simulations for the bulk crystals, our
understanding of semiconductor NCs remains undeveloped. Traditional
\emph{ab initio} density functional theory (DFT) calculations have
been successfully applied to explore the geometry and
\emph{electronic} structure of small covalent semiconductor NCs, and
rarely, up to experimental sizes \cite{zhou:066704}. However, accurate
calculation of the absorption spectrum of NCs requires methods
\emph{beyond} DFT that are usually computationally prohibitive. While
DFT is a ground state theory, it is usually assumed that one can use
the DFT electronic gap \cite{note6} with some ``rigid shift'' to
calculate the absorption treshold of the NCs. For example, a
difference of $\approx$1.5~eV was deduced by comparing the DFT and
Quantum Monte Carlo results on ideal very small hydrogenated silicon
nanocrystals \cite{PhysRevLett.88.097401}. However, it is not clear
whether this type of correction may hold for the reconstructed
surface. In addition, the surface states may modify not just the
absorption treshold of the ideal nanocrystals, but the other part of
the spectrum as well.

In this Letter we investigate ideal and reconstructed hydrogenated
cubic silicon carbide nanocrystals (SiCNC). The choice of this
material was made for several reasons: i) as a binary compound
material one can study the effect of two types of termination ii)
SiCNC is one of the most promising candidates for bioinert biomarkers
\cite{note2,botsoa:173902,fan:1058} and environmental friendly
nanooptics \cite{ReboredoF.A._nl049876k,makkai:253109,peng:024304}
iii) the experimental absorption spectrum is available for
small-medium sized colloidal SiCNCs \cite{botsoa:173902}. We
determined the structure by standard DFT calculations (PBE) while the
absorption spectrum was calculated by hybrid functional based
time-dependent DFT (TDPBE0) calculations. We found that i) the TDPBE0
absorption treshold of the ideal SiCNCs is higher than their PBE
electronic gap by approximately 1~eV ii) the reconstructed surface can
significantly change not just the onset of absorption, but the
\emph{shape} of the spectrum at higher energies compared to that of
the ideal SiCNCs. Our calculations indicate that not just the position
of the absorption treshold but the shape of the absorption spectrum
can experimentally discriminate the ``ideal'', and the reconstructed,
hydrogenated covalent NCs.

Model NCs were constructed by a cutting approximately spherical
clusters out of the perfect cubic SiC crystal in such a way that
maximum two dangling bonds per surface atoms were allowed. The radii
of the resulting clusters are ranged from 0.9 to 2.8~nm. The
spherical shape was chosen based on the results of transmission
electron microscope measurements (TEM) results
\cite{wu:026102,botsoa:173902,fan:1058}. The dangling bonds of the
surface atoms were saturated by hydrogen atoms. The geometry of these
structures were optimized by using the DFT PBE functional
\cite{PBE}. We will refer to these relaxed NCs as the ``ideal'' ones.
%
%
Further relaxation after removing some hydrogen atoms allows the
formation of long bonds on the surface. Typically, 2$\times$1-type
reconstruction occurs on the (100) facets but other reconstructions
were found as well. For geometry optimization we have utilized the
\textsc{SIESTA} code \cite{0953-8984-14-11-302} using a double-$\zeta$
polarized basis set and Troullier-Martins pseudopotentials
\cite{Troullier91}. Since \textsc{SIESTA} is a supercell code, at
least a distance of 10~\AA\ was ensured between periodic images. All
the atoms were allowed to relax until the forces were below
0.02~eV/\AA.  The $d$$\approx$1.0~nm SiCNCs were studied by well
converged plane wave calculations \cite{note3}. The results agreed
within 0.01~\AA\ for the geometries and 0.1~eV for the electronic
gaps. Due to the self-interaction error, PBE underestimates the
electronic gap, and the best remedy for this problem would be the use
of a GW-method \cite{hedin:1969}. However, GW-method would impose
strict limits on the NC size. Instead, a hybrid functional may be
applied, since those were shown to reproduce the experimental gap
quite well for many different systems \cite{marsman:064201}. We
applied the PBE0 functional to calculate the electronic structure at
the PBE geometry \cite{perdew:9982}. While the electronic gap obtained
this way could be rather accurate, the large electron-hole interaction
in SiCNCs \cite{PhysRevB.68.035334} can reduce the observed
``optical'' gap in the absorption spectrum considerably.  Therefore,
we applied time-dependent DFT (TDDFT) with the PBE0 functional in the
kernel (TDPBE0), in order to obtain the absorption spectrum of the
SiCNCs. We define the exciton binding energy as the difference of the
PBE0 electronic gap and the optical gap. \emph{We were able to
  calculate the absorption spectrum for NCs up to $d$$\approx$1.5~nm ,
  i.e, the experimental size range could be reached.}  For these
calculations the \textsc{Turbomole} cluster code
\cite{bauernschmitt:454} was utilized, applying an all-electron
Gaussian basis set with the same quality for the valence electrons as
used in the \textsc{SIESTA} code. We used a frequency-domain
TDDFT-method, that allowed us to span the selected region of the
excitation energy and analyze the nature of the transition
\cite{bauernschmitt:454}. In particular, we determined the
application-relevant low energy absorption spectrum for which
experimental data are available \cite{botsoa:173902}. Similar hybrid
functional based TDDFT-methods yielded superior results for small
hydrogen nanoclusters and molecules over TDPBE results, and fairly
reproduced the experimental spectrum and/or highly accurate quantum
chemistry results \cite{2006PhRvB..74d5433L}.
\begin{figure}
\includegraphics[keepaspectratio,width=8cm]{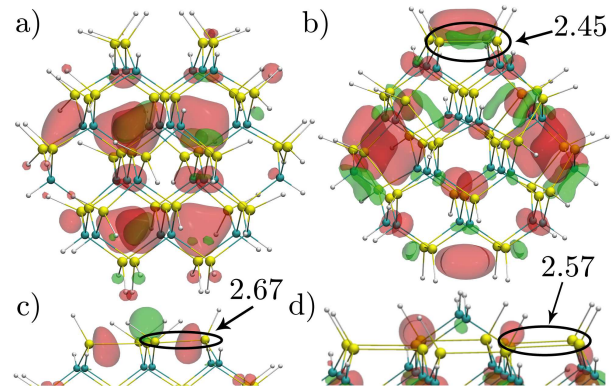}
  \caption{\label{fig:SiCNC}(Color online) a) The ``ideal'' $d$=1.0~nm
    SiCNC, b) typical Si 2$\times$1-type reconstruction in $d$=1.0~nm
    SiCNC c) a part of 3$\times$1-type reconstructions in $d$=1.5~nm SiCNC d)
    a part of step-like reconstruction in $d$=2.1~nm SiCNC.
    The bond length of the long Si-Si bonds are shown in \AA . 
    Small(big) cyan(yellow) and white balls represent C(Si) and H
    atoms, respectively. The red(green) lobes show the
    positive(negative) isosurfaces of the LUMO states.
    }
\end{figure}


We first considered the ideal SiCNCs. There are two possibilities to
construct a spherical NC at any size: a C- and a Si-atom in the
center. For instance, an NC with $d$$\approx$0.9~nm contains
altogether 35 atoms. If Si is at the center, this means 19 Si and 16 C
atoms, if C is at the center then the atom counts are reversed.  For
$d$$\leq$1.4~nm, these two types of SiNCs yield considerable
difference in both the electronic and optical gaps (see
Table~\ref{tab:gap}). Generally, the gaps are smaller for those NCs
that contain more Si-H than C-H bonds at the surface. For larger NCs
the difference is below $\approx$0.2-0.1~eV.

As apparent from Fig.~\ref{fig:SiCNC}a, there may be strong steric
repulsion between some H-atoms on the surface of the ideal NCs. By
removing such close H pairs, we allowed the surface to reconstruct
while keeping the stoichiometry of the SiCNCs. The steric repulsion is
especially strong for the Si-terminated surface. This is expected
since the second neighbor distance between the Si-atom is about
3.0~\AA\ in SiC, and the corresponding Si-H bonds are about 1.5~\AA
. This effect is much less pronounced for the C-terminated surface
where the C-H distance is about 1.1~\AA . Typically, 2$\times$1-type
reconstructions occurred at the (100) facets (see
Fig.~\ref{fig:SiCNC}b), in agreement with previous PBE calculations
\cite{ReboredoF.A._nl049876k}. In some cases, we have found
3$\times$1-type (Fig.~\ref{fig:SiCNC}c) and step-like
(Fig.~\ref{fig:SiCNC}d) reconstructions at Si-terminated facets. We
note that the Si--Si bonds are longer in 3$\times$1-type and step-like
reconstructions than in 2$\times$1-type reconstructions. The
appearance of the 3$\times$1-type and step-like reconstructions
depends strongly on the actual shape of SiCNC surface, and are not as
common as the 2$\times$1-type reconstructions at the Si-terminated
(100) facets. 
%
%
If the formation of SiCNCs is driven by kinetics, then in principle
any of the surfaces previously discussed could be formed. If SiCNCs
are formed at near equilibrium conditions, then 2$\times$1-type
reconstructions could occur most frequently beside the ideal ones
\cite{ReboredoF.A._nl049876k}.

\begin{figure}
\includegraphics[keepaspectratio,width=6.5cm]{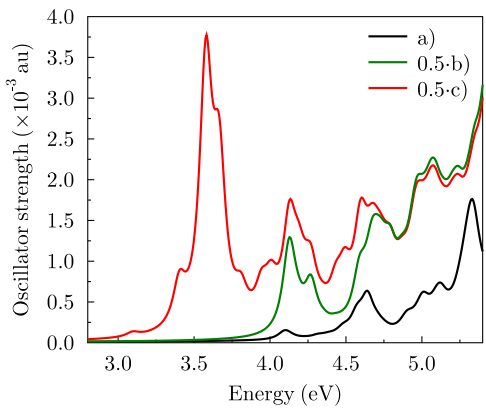}
\caption{\label{fig:spectrum} (Color online) The calculated absorption
  spectrum of the ensemble of 1.0$\leq d$$\leq$1.5~nm SiCNCs a)
  without b) with 2$\times$1-type reconstructed configurations. c) the
  same with b) plus 3$\times$1-type reconstruction. b) and c) are
  scaled by 0.5 as indicated.}
\end{figure}

The calculated gaps of the reconstructed SiCNCs can be
found in Table~\ref{tab:gap}.
\begin{table*}
  \caption{\label{tab:gap} The atomic structure of the studied ideal
    and reconstructed SiCNCs and their calculated gaps. First, second
    and third columns: number of all atoms (\#All) with labeling the
    reconstruction type in parentheses, total number of C and Si atoms
    (\#C,Si), and its diameter ($d$) in nm. Fourth, fifth and sixth
    columns: PBE, PBE0 and TDPBE0 gaps in eV,
    respectively. H$\rightarrow$L: the contribution of
    HOMO$\rightarrow$LUMO transition to the absorption onset in \%.
    $\Delta_\text{QP}$=PBE0$-$PBE gaps, i.e., the ``quasi-particle''
    corrections. $\Delta_\text{Exc}$=TDPBE0$-$PBE0 gaps, i.e., the
    binding energy of the exciton. $\Delta_\text{Corr}$=TDPBE0$-$PBE
    gaps. The $A/B$ means the values for Si/C centered SiCNCs,
    respectively.}
\begin{ruledtabular}
\begin{tabular}{rcclllllll}
 \#All  &  \#C,Si & $d$(nm)  &   PBE(eV)   &  PBE0(eV)  &  TDPBE0(eV)
 & H$\rightarrow$L(\%) &   $\Delta_\text{QP}$(eV) &
 $\Delta_\text{Exc}$(eV)  &  $\Delta_\text{Corr}$(eV) \\
\hline
71            &  35 & 0.9 & 4.61/4.85 & 6.48/6.75 & 5.53/5.83 &
97/80 & 1.87/1.90 & -0.95/-0.92 & 0.92/0.98 \\
 130            &  66 & 1.0 & 3.79/4.09 & 5.61/5.91 & 4.88/5.03 &
 97/97 & 1.82/1.82 & -0.73/-0.88 & 1.09/0.94 \\
 106(2$\times$1) &  66 & 1.0 & 3.20/3.17 & 4.87/4.94 & 4.27/4.13 &
 0/64 & 1.67/1.77 & -0.60/-0.81 & 1.07/0.96 \\
 142            &  78 & 1.1 & 3.81/3.90 & 5.59/5.69 & 4.75/4.89 &
 97/94 & 1.78/1.79 & -0.84/-0.81 & 0.94/0.99 \\
 130(2$\times$1) &  78 & 1.1 & 3.54/3.57 & 5.33/5.28 & 4.59/4.69 & 
 94/0  & 1.79/1.71 & -0.74/-0.58 & 1.06/1.13 \\
 163            &  87 & 1.2 & 3.66/3.58 & 5.44/5.35 & 4.64/4.57 &
 97/94 & 1.78/1.77 & -0.80/-0.79 & 0.98/0.99 \\
 247            & 147 & 1.4 & 3.24/3.30 & 4.96/5.03 & 4.22/4.31 &
 96/84 & 1.72/1.73 & -0.74/-0.72 & 0.98/1.01 \\
 223(2$\times$1) & 147 & 1.4 & 3.11/2.82 & 4.77/4.55 & 4.19/4.18 & 
 95/0 & 1.66/1.73 & -0.58/-0.37 & 1.08/1.37 \\
 292            & 172 & 1.5 & 3.11/3.06 & 4.82/4.78 & 4.11/4.08 &
 96/93   & 1.71/1.72 & -0.71/-0.69 & 1.00/1.01 \\
 292(3$\times$1) & 172 & 1.5 & 1.96/-    & 3.51/-    & 3.10/-    &   
 0/-  & 1.55/-    & -0.41/-     & 1.14/-    \\
\end{tabular}
\end{ruledtabular}
\end{table*}
We also give the calculated TDPBE0 absorption spectrum in
Fig.~\ref{fig:spectrum}, averaged for all the considered SiCNCs by
excluding or including the reconstructed NCs. We show the effect of
the typical 2$\times$1-type (Fig.~\ref{fig:spectrum}b) and the
atypical 3$\times$1-type reconstructions (Fig.~\ref{fig:spectrum}c)
separately. We used Lorentzian broading of 0.05~eV for each individual
peak to construct the spectrum.

Several very important issues can be concluded from
Table~\ref{tab:gap} and Fig.~\ref{fig:spectrum}. We first analyze the
ideal SiCNCs. In the given size range of the SiCNCs, the electronic
gap obtained with PBE0 is about 1.9-1.7~eV larger than the one from
PBE, while the difference in optical gaps is about half as large. It
should be noted that that both the PBE0$-$PBE difference
(corresponding roughly to the necessary quasiparticle correction) and
the TDPBE0$-$PBE0 difference (the binding energy of the exciton) are
decreasing continuously with increasing size by approximately the same
amount (by about 0.3~eV from $d$$\approx$0.9~nm to
$d$$\approx$1.5~nm). Both effects are understandable, if we consider
that by increasing the size of the NCs, the screening is enhanced,
reducing the quasi-particle correction, while the HOMO and LUMO wave
functions can become more delocalized, increasing the exciton
radius. Since the two effects cancel each other almost perfectly,
\emph{we find that by adding 1.0~eV to the PBE gaps we can reproduce
  the calculated TDPBE0 optical gaps within 0.09~eV for
  0.9$\leq$$d$$\leq$1.5~nm SiCNCs.} We note that the first absorption
peak is primarily the result of the transition between the HOMO and
LUMO states, and therefore the ``rigid shift'' correction is justified.


The 2$\times$1-type reconstruction causes a large reduction in the
optical gap of the $d$$\approx$1.0~nm SiCNCs compared to their ideal
counterparts, while a small red-shift ($\approx$0.2~eV) appears for
the larger SiCNCs. This red-shift is associated with the reduced
number of Si-H bonds. These findings can be understood, considering
that the longer surface Si--Si bonds lead to a smaller
bonding-antibonding splitting, and to the appearance of localized
surface states in the ``bulk'' gap of the ideal small SiCNC. As the NC
size increases and the bulk gap decreases, these surface states merge
with the ``bulk''-like states of ideal SiCNCs. For example, the
smallest optical gap (3.10~eV), is found for an atypical
3$\times$1-type reconstruction with a very long Si--Si bond.  With a
single exception, the PBE+1.0~eV$\approx$TDPBE0 ``rule'' seems to
apply for the reconstructed surfaces, too (see $\Delta _\text{Corr}$
in Table~\ref{tab:gap}). However, this is a fortuitious cancellation
of errors, since the first absorption peak often does \emph{not} come
from the HOMO-LUMO transition: it can be absolutely dark
(H$\rightarrow$L=0 in Table~\ref{tab:gap}). Generally, these surface
states must be studied one-by-one, and one should not assume a very
simple rigid shift correction.
%
The calculated absorption spectrum of the ideal SiCNCs shows an
overall monotone increasing function in the given range (see
Fig.~\ref{fig:spectrum}). We found a typical peak at about 4.6~eV due
to the wave functions that cause the $\Gamma \rightarrow L$ enhanced
absorption at similar energy in bulk cubic SiC. As apparent from
Fig.~\ref{fig:spectrum}, the reconstructed surface \emph{alters not
  just the absorption treshold but the shape of the spectrum} compared
to the ideal NCs. The huge peak at about 3.5~eV comes primarily from
3$\times$1-type reconstructions. The overall larger absorption from
the reconstructed surfaces is partly due to the enhanced dipole
transition strength between the localized states, and partly due to
higher density of states near the gap. For these small/medium SiCNCs
the quantum confinement effect can be fully obscured by reconstructed
SiCNCs.

\begin{figure}
\includegraphics[keepaspectratio,width=6.5cm]{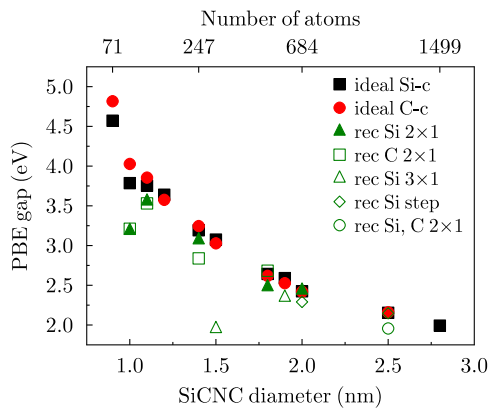}
  \caption{\label{fig:idealrec}(Color online) 
   The calculated PBE gap of the ideal and reconstructed SiCNCs. Si-c(C-c)
   means Si(C) atoms in the center.
    }
\end{figure}

We report the PBE electrical gap for large SiCNCs in
Fig.~\ref{fig:idealrec}. Here, we allowed the atypical 3$\times$1-type
or step-like reconstructions together with the typical 2$\times$1-type
reconstruction in the same SiCNC. The trend indicates that the states
associated with the atypical reconstructions will merge with the
``bulk''-like states of the ideal $d$$\geq$3~nm SiCNCs. Therefore, the
surface reconstruction does not change the absorption treshold of the
ideal counterparts. For the step-like atypical reconstructions
(Fig.~\ref{fig:SiCNC}d) we found an enhanced density of states close
to HOMO level that may result in somewhat enhanced absorption compared to
their ideal counterpart. Nevertheless, the surface to volume ratio
grows rapidly with the increasing size of the SiCNCs. Therefore, we
predict that the states from surface reconstructions cannot be
recognized for $d$$\geq$3~nm SiCNCs.
We note that Botsoa \emph{et al.} found the first absorption peak at
1.8~eV for colloidal SiCNCs where $d$$\approx$1.5~nm SiCNCs were in
the vast majority \cite{botsoa:173902}. Another group measured this
onset to occur at $\approx$2.75~eV for colloidal
$d$$\approx$3.9$\pm$1.1~nm SiCNCs \cite{fan:1058}. This runs against
the expected trend of the quantum confinement effect
\cite{PhysRevB.68.035334} that may be explained by surface
reconstructions. However, the onset and the \emph{shape} of our calculated
absorption spectrum do not show the features of Botsoa's experimental
data (c.f., Fig.~2a in Ref.~\onlinecite{botsoa:173902} and our
Figs.~\ref{fig:spectrum}b,c). We could explain those features by oxygen
defects at the surface \cite{Voros09}.

In summary, we determined the optical gaps of hydrogenated silicon
carbide nanocrystals. We found that the surface reconstruction can
play a crucial role for small SiCNCs, and we showed that the onset and
the \emph{shape} of the absorption spectrum provide useful method for
discriminating between reconstructed and ideal nanocrystals.

AG acknowledges the support from Hungarian OTKA No.\ K-67886 and the
NIIF Supercomputer center grant No.~1090 while MV the support from GE
Lighting Hungary. The bilateral support from MTA-DFG No.~194 is
acknowledged. The fruitful discussions with Thomas Niehaus and Jan
Knaup are appreciated.


\begin{thebibliography}{10}

\bibitem{Michalet05}
X. Michalet, F.~F. Pinaud, L.~A. Bentolila, J.~M. Tsay, S. Doose, J.~J. Li, G.
  Sundaresan, A.~M. Wu, S.~S. Gambhir, and S. Weiss, Science {\bf 307},  538
  (2005).

\bibitem{Gur05}
I. Gur, N.~A. Fromer, M.~L. Geier, and A.~P. Alivisatos, Science {\bf 310},
  462  (2005).

\bibitem{Klimov07}
V.~I. Klimov, S.~A. Ivanov, J. Nanda, M. Achermann, I. Bezel, J.~A. McGuire,
  and A. Piryatinski, Nature {\bf 447},  441  (2007).

\bibitem{Pandey08}
A. Pandey and P. Guyot-Sionnest, Science {\bf 322},  929  (2008).

\bibitem{zhou:066704}
Y. Zhou, Y. Saad, M.~L. Tiago, and J.~R. Chelikowsky, Physical Review E {\bf
  74},  066704  (2006).

\bibitem{note6}
The electronic gap is defined as the difference of the single particle levels
  of the highest occupied molecular orbital (HOMO) and the lowest unoccupied
  molecular orbital (LUMO).

\bibitem{PhysRevLett.88.097401}
A. Puzder, A.~J. Williamson, J.~C. Grossman, and G. Galli, Phys. Rev. Lett.
  {\bf 88},  097401  (2002).

\bibitem{note2}
Adam Gali's seminar at the Physics Department of Harvard University (2007,
  May).

\bibitem{botsoa:173902}
J. Botsoa, V. Lysenko, A. G\'{e}lo\"{e}n, O. Marty, J.~M. Bluet, and G.
  Guillot, Applied Physics Letters {\bf 92},  173902  (2008).

\bibitem{fan:1058}
J. Fan, H. Li, J. Jiang, L.~K.~Y. So, Y.~W. Lam, and P.~K. Chu, Small {\bf 4},
  1058  (2008).

\bibitem{ReboredoF.A._nl049876k}
F. Reboredo, L. Pizzagalli, and G. Galli, Nano Letters {\bf 4},  801  (2004).

\bibitem{makkai:253109}
Z. Makkai, B. P\'{e}cz, I. B\'{a}rsony, G. Vida, A. Pongr\'{a}cz, K.~V.
  Josepovits, and P. De\'{a}k, Applied Physics Letters {\bf 86},  253109
  (2005).

\bibitem{peng:024304}
X.-H. Peng, S.~K. Nayak, A. Alizadeh, K.~K. Varanasi, N. Bhate, L.~B. Rowland,
  and S.~K. Kumar, Journal of Applied Physics {\bf 102},  024304  (2007).

\bibitem{wu:026102}
X.~L. Wu, J.~Y. Fan, T. Qiu, X. Yang, G.~G. Siu, and P.~K. Chu, Physical Review
  Letters {\bf 94},  026102  (2005).

\bibitem{PBE}
J.~P. Perdew, K. Burke, and M. Ernzerhof, Phys. Rev. Lett. {\bf 77},  3865
  (1996).

\bibitem{0953-8984-14-11-302}
J.~M. Soler, E. Artacho, J.~D. Gale, A. Garc\'{i}a, J. Junquera, P.
  Ordej\'{o}n, and D. S\'{a}nchez-Portal, Journal of Physics: Condensed Matter
  {\bf 14},  2745  (2002).

\bibitem{Troullier91}
N. Troullier and J.~L. Martins, Phys. Rev. B {\bf 43},  1993  (1991).

\bibitem{note3}
For this purpose the \textsc{PWSCF} code (P. Giannozzi \emph{et al.},
  http://www.quantum-espresso.org) was applied with using standard ultrasoft
  pseudopotentials and 40~Ry plane wave cut-off.

\bibitem{hedin:1969}
L. Hedin and S. Lundqvist,  in {\em Solid State Physics}, edited by H.
  Ehrenreich, F. Seitz, and D. Turnbull (Academic, New York, 1969), Vol.~23.

\bibitem{marsman:064201}
M. Marsman, J. Paier, A. Stroppa, and G. Kresse, Journal of Physics: Condensed
  Matter {\bf 20},  064201 (9pp)  (2008).

\bibitem{perdew:9982}
J.~P. Perdew, M. Ernzerhof, and K. Burke, Journal of Chemical Physics {\bf
  105},  9982  (1996).

\bibitem{PhysRevB.68.035334}
D.~H. Feng, Z.~Z. Xu, T.~Q. Jia, X.~X. Li, and S.~Q. Gong, Phys. Rev. B {\bf
  68},  035334  (2003).

\bibitem{bauernschmitt:454}
R. Bauernschmitt and R. Ahlrichs, Chem. Phys. Letters {\bf 256},  454  (1996).

\bibitem{2006PhRvB..74d5433L}
O. Lehtonen and D. Sundholm, Physical Review B {\bf 74},  045433  (2006).

\bibitem{Voros09}
M. V\"or\"os, P. De\'ak, T. Frauenheim, and A. Gali, unpublished.

\end{thebibliography}

\end{document}